\documentclass[
prl,
twocolumn,
superscriptaddress,
nolongbibliography,
 amsmath,amssymb,
 aps,
]{revtex4-2}

\usepackage[linkcolor = blue, citecolor = blue, urlcolor = blue, colorlinks = true]{hyperref}
\usepackage[dvipsnames]{xcolor}
\usepackage{graphicx}
\usepackage{dcolumn}
\usepackage{bm}

\begin{document}

\title{Multiscale order, flocking and phenotypic hysteresis\\in the cellular Potts model of epithelia}

\author{Calvin C. Bakker}
\affiliation{Instituut-Lorentz, Leiden Institute of Physics, Universiteit Leiden, P.O. Box 9506, 2300 RA Leiden, The Netherlands}
\author{Marc Durand}
\affiliation{Universit\'{e} Paris Cit\'{e}, CNRS, Mati\`{e}re et Syst\`{e}mes Complexes, F-75013 Paris, France}
\author{Fran\c{c}ois Graner}
\affiliation{Universit\'{e} Paris Cit\'{e}, CNRS, Mati\`{e}re et Syst\`{e}mes Complexes, F-75013 Paris, France}
\author{Luca Giomi}
\affiliation{Instituut-Lorentz, Leiden Institute of Physics, Universiteit Leiden, P.O. Box 9506, 2300 RA Leiden, The Netherlands}
\email{giomi@lorentz.leidenuniv.nl}

\begin{abstract}
In epithelia, how do collective cell migration and tissue spatial organization feedback on each other? We address this question through large-scale numerical simulations of the cellular Potts model. By accounting for both cell morphology and cytoskeletal activity, we uncover a remarkably rich phase diagram featuring multiple types of orientational order, either as distinct phases or coexisting across length scales. We identify a specific pathway in parameter space along which a gradual increase in the actin polymerization rate drives a phase transition into a long-range flocking state. Simultaneously, quasi-long-range nematic order emerges at length scales much larger than the cell size due to the combined effects of directed motion and lateral cell-cell interactions. At length scales comparible to cell size, however, cells adopt an approximatively hexagonal morphology, resulting in hexanematic order, similar to that observed in reconstituted Madin-Darby Canine Kidney (MDCK) cell monolayers. With further increases in actin polymerization, nematic order becomes fully long-range, while hexatic order remains quasi-long-range and confined to short length scales, but independent of cytoskeletal activity. When noise is sufficiently low to allow crystallization at finite actin polymerization rate, cycling the cell-monolayer across the melting transition yields an example of phenotypical hysteresis, reminiscent of that observed across the epithelial-mesenchymal transition.
\end{abstract}

\maketitle

The attempt of deciphering ordered structures in the seemingly structureless organization of epithelial tissues has recently led to the notion of {\em multiscale order} (MsO): i.e. the coexistence of different non-equilibrium phases of matter across length scales~\cite{armengol2023epithelia,eckert2023hexanematic}. While this program inevitably requires extending the conventional definition of ``phase'', as inherited from equilibrium thermodynamics, recent numerical and experimental studies have progressively indicated that aspects of this complexity can be captured using the powerful language of liquid crystal hydrodynamics, with thermal equilibrium marking one end of a potentially large spectrum of novel out-of-equilibrium behaviors~\cite{armengol2024hydrodynamics,carenza2025quasi}. The role of hexatic order, in particular, has drawn the most attention by virtue of its manifest connection with the honeycomb structure of cellular networks~\cite{li2018role,durand2019thermally,pasupalak2020hexatic,li2021melting,nemati2024cellular}. 

Hexatic liquid crystals are two-dimensional fluids characterized by quasi long-range $6$-fold orientational order, generally quantified in terms of the complex function $\psi_{6}=e^{6i\vartheta}$, with $\vartheta$ the local orientation of building blocks. Unlike in long-range ordered systems, where orientational correlations remains finite even at an infinite distance, quasi long-range order (QLRO) implies an algebraically decaying orientational correlation function -- i.e. $\langle \psi_{6}(\bm{r})\psi_{6}^{*}(\bm{0}) \rangle \sim |\bm{r}|^{-\eta_{6}}$ -- and a scale-dependent order parameter: i.e. $\Psi_{6}(\ell)=\langle \psi_{6} \rangle_{\ell} \sim \ell^{-\eta_{6}/2}$, where $\langle \cdots \rangle_{\ell}$ denotes an ensemble average over the length scale $\ell$ and $\eta_{6}$ a non-universal exponent (or anomalous dimension). At equilibrium, this exponent varies in the range $0 < \eta_{6} \le 1/4$, with the lower bound corresponding to long-ranged order (LRO) and the upper bound setting the threshold above which a homogeneous hexatic phase is unstable to the unbinding of $5$--$7$ {\em dislocations} into pairs of $5$- and $7$-fold {\em disclinations}~\cite{halperin1978theory,nelson1979dislocation,young1979melting}. In epithelia, this process occurs through the proliferation of topological rearrangements of the first kind (i.e. T1), thus leading to cell intercalation and collective but {\em undirected} migration (i.e. swarming). In this regime, $1/4 < \eta_{6}\le 2$, with the upper bound corresponding an ensemble of uniformly distributed random orientations. Furthermore, well in the migrating phase, the interplay between lateral interactions and motion results into the emergence of nematic order at large length scales. By contrast, hexatic order remains dominant over intermediate length scales that are typically one to two orders of magnitude above the average cell size. This particular example of multiscale organization, named {\em hexanematic} in Ref.~\cite{armengol2023epithelia}, was observed in reconstituted layers of MDCK cells, as well as in numerical simulations of the multi phase-field model~\cite{armengol2023epithelia,eckert2023hexanematic}. Yet, how nematic order develops at the large scale and what hexanematic organization entails in practice is presently unknown. 

On the other hand, large-scale {\em directed} migration (i.e. flocking) appears prominent in fluidized layers of epithelial cells, both in bulk and under confinement~\cite{carmona2008contact,doxzen2013emergent,malinverno2017edocytic,giavazzi2017giant,glentis2022emergence}. This ability of moving collectively and coherently in space is instrumental to a large variety of developmental process, such as gastrulation~\cite{scarpa2016collective} and neural crest migration~\cite{szabo2018mechanism}, as well as in cancer invasion~\cite{carrillo2022spatiotemporal,argento2025three}. In these processes, the typical scale of collectively moving structures ranges from a dozen of border cells, as in the development of the egg chamber in {\em Drosophila}, to several hundred thousand cells, as in {\em Dictyostelium} slug development~\cite{weijer2009collective}. Now, because of its resemblance with flocking in active systems, various active matter models -- i.e. generically inspired by Vicsek's foundational work~\cite{vicsek1995novel} -- have been used to rationalize the origin of large-scale directed migration in epithelia (see e.g. Refs.~\cite{giavazzi2017giant,giavazzi2018flocking}). In these models, flocking is recovered in the presence of {\em self-alignment} --  i.e. the property of polar active particles to align their velocity with respect to their polar axis (see Ref.~\cite{baconnier2025self} for a recent review) -- while the role of orientational order is generally ignored. The latter rises questions on how flocking occurs in tissues, where self-alignment is itself an emergent property rooted in more fundamental physical mechanisms and biochemical pathways, and how these two distinct classes of migratory patterns -- i.e. directed and flocking-based or undirected and intercalation-based -- alternate along the epithelial-mesenchymal spectrum~\cite{zhang2018epithelial}.

In this article we address these questions, and more broadly the feedback between collective cell migration and spatial cell organization~\cite{yu2025feedback}, via large-scale cell-based numerical simulations of the cellular Potts model (CPM)~\cite{graner1992simulation,glazier1993simulation}. Combining a recent implementation of the CPM on Graphical Processing Units (GPUs)~\cite{sultan2023parallelized} with an explicit model of the actin cytoskeleton~\cite{niculescu2015crawling}, allows us to map out the complete phase diagram of epithelial migration {\em in silico} and correlate the resulting migratory patterns -- comprising both directed and undirected motion -- with two distinct manifestations of MsO. 

\begin{figure}
\includegraphics[width=\columnwidth]{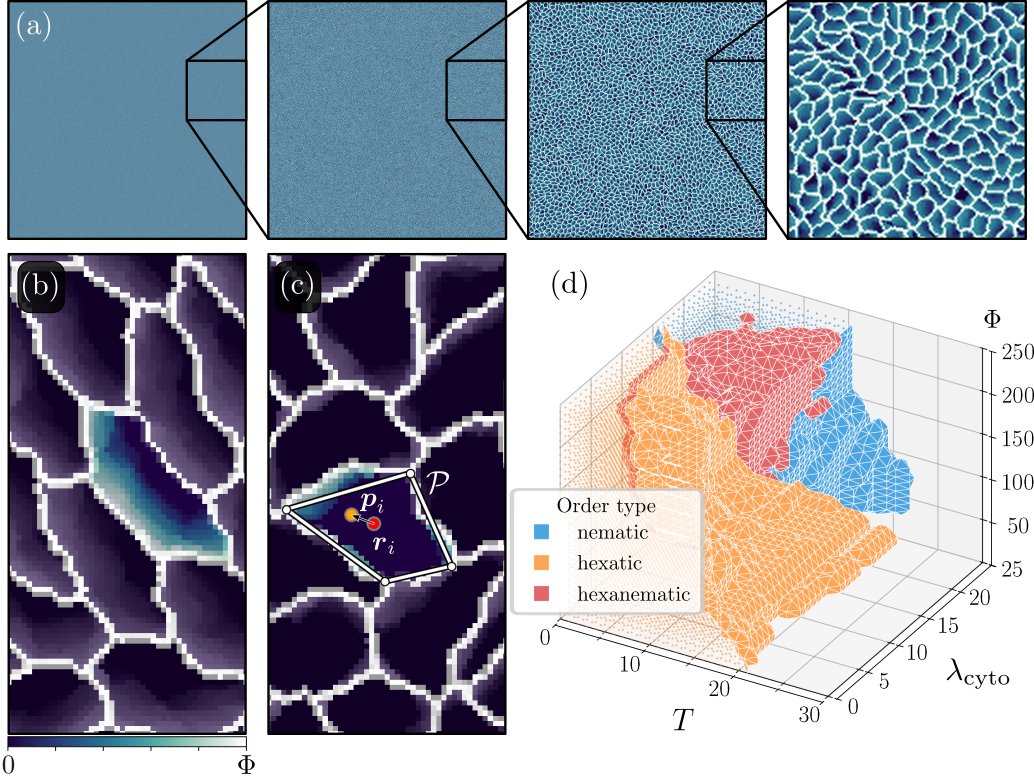}
\caption{
\label{fig:figure1}
An overview of the phases of matter produced by the cellular Potts model. (a) CPM configuration of $\approx7\times 10^5$ cells on a $(8192,8192)$ pixel grid, zoomed in thrice to $1/64$'th of the grid to display the scale of the simulations performed. (b) Typical configuration of tread-milling cell corresponding to large $\Phi$ values. (c) For weak cytoskeletal activity, cells are more isotropic and multiple protrusions form and retract. (d) Phase diagram obtained by varying the parameters $\lambda_{\rm cyto}$, $\Phi$ and $T$; obtained by the RDFCA algorithm where boundaries are shown of the classified phases. Blue, orange, and red boundaries indicate the dominance of nematic, hexatic, and hexanematic order in the monolayer respectively.}
\end{figure}

In the CPM, cells are represented as connected sets of pixels in a two-dimensional $L\times L$ array (Fig.~\ref{fig:figure1}a--c). 
Each pixel stores two integer-valued fields. One is the cell label number, $\sigma_{i}\in \{1,\,2\ldots\,N\}$, with $N$ the total number of cells: a cell is defined as the set of pixels which share a common label. The other, $\phi_{i}\in\{1,\,2\ldots \Phi\}$, with $\Phi$ a pre-defined constant. The former assigns a common label to the pixels comprising a given cell, while the latter, introduced in Ref.~\cite{niculescu2015crawling} as part of the so-called Act-model, reflects the density of the actin filaments forming the cell's lamellipodium (Fig.~\ref{fig:figure1}b,c). The dynamics is generated via Markov chain Monte Carlo method based on the Hamiltonian
$
\mathcal{H}
=J\sum_{\langle ij \rangle}(1-\delta_{\sigma_{i}\sigma_{j}})
+\lambda_{\rm cell}\sum_{c}(A_{c}-A_{0})^{2} 
+\lambda_{\rm cyto}\sum_{i}{\rm GM}\left(\phi_{i}/\Phi\right)
$%
~\cite{wortel2021local,van2022computational}. The first term, where the summation extended to all pairs of neighboring pixels, embodies the lateral surface interactions among cells, while the second term sets the constraint on of the cellular area $A_{c}$. The last term governs the assembly of actin filaments, with ${\rm GM}(x_{i})=(\prod_{j \in M_{i}}x_{j})^{1/\Vert M_{i}\Vert}$, the geometric mean over the Moore neighborhood of the pixel~\cite{toffoli1987cellular}. At each Monte Carlo step (MCS), the array is updated via the Metropolis-Hastings algorithm, with nominal fluctuation allowance $T$ (analogous to an ``effective temperature'')~\footnote{A MCS here is equal to an amount of update-attempts of the Metropolis-Hastings algorithm that is equal to the amount of pixels in the two-dimensional $L \times L$ array: $L^2$.}. As a pixel is incorporated into a new cell, its corresponding $\phi_{i}$ value is initialized to $\Phi$ and then progressively decreased by one unit after each of the next iterations: i.e. $\phi_{i}(t+1)=\phi_{i}(t)-1$. The persistence entailed in this process, combined with the spatiotemporal dynamics governed by Hamiltonian,  favors a translation of the pixels towards regions of large $\phi_{i}$ value, giving rise to tread-milling (see Fig.~\ref{fig:figure1}b). The constants $J$, $A_{0}$, $\lambda_{\rm cyto}$, $\lambda_{\rm cell}$, $\Phi$ and $T$ together constitute the set of material parameters of the model. Finally we use the algorithm introduced in Ref.~\cite{durand2016efficient} to restrict the space of possible pixel rearrangements to prevent cell fragmentation and to achieve confluency in the {\em in silico} epithelial layers~\cite{SI}.

\begin{figure*}[t]
\includegraphics[width=\textwidth]{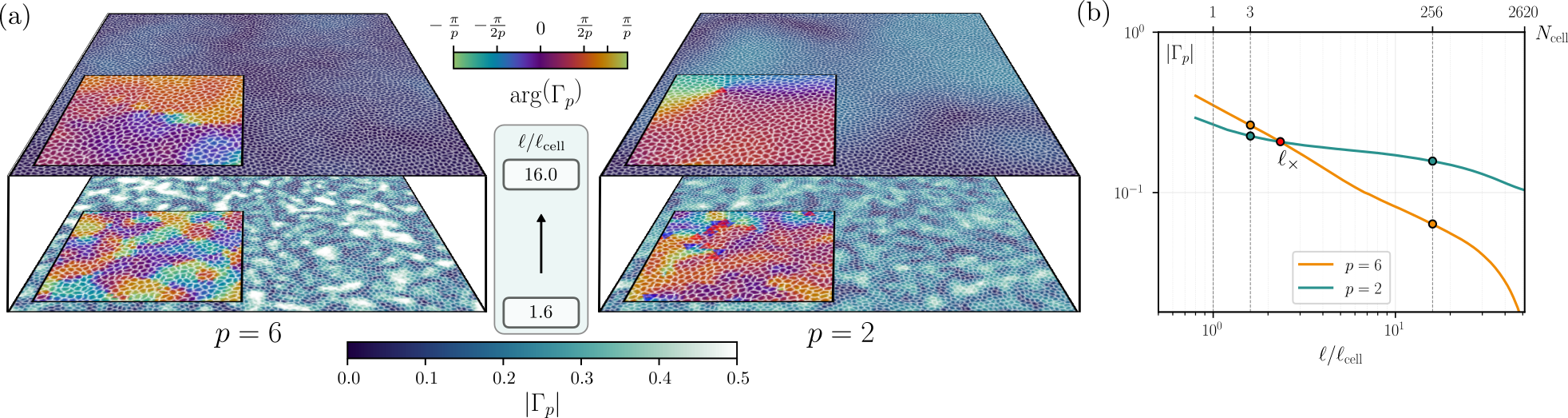}
\caption{
    \label{fig:figure2} 
    The cellular Potts model (CPM) in a hexanematic state. In (a) a single snapshot of the CPM is shown by the cell outline, with cells colored by the magnitude of the order parameter $|\Gamma_p|$ coarse-grained over a length scale of $\ell$. Here the stacked left and right panels display $\Gamma_6$ and $\Gamma_2$ respectively. In the sub-panel the value of $\arg(\Gamma_p)$ is shown for a subsection of the full grid. Note that the colors of this sub-panel are dependent on the orientational order parameter $p$. 
    In (b) the coarse grained $\Gamma_p$ curves are shown, with the hexanematic crossover point $\ell_\times$ shown with a red circle, and the values attributed to the four panels in (a) given by orange and blue circles.
}
\end{figure*}

In all our simulations, we set $A_0=96.51$ and $J=16.0$~\cite{durand2016efficient}. We focus on the effects of cytoskeletal activity and fluctuations by varying $T$ and the parameters $\lambda_{\rm cyto}$ and $\Phi$. To efficiently sample this three-dimensional parameter space, we randomly generate $901$ distinct parameter configurations, each of which is used to perform a numerical simulation consisting of $10^6$ MCS on grids of size $L = 2^{10}$, which we found to be sufficient to answer the questions under consideration. Our simulations are performed on NVIDIA T600 and RTX 2080ti graphical processing units following the implementation by Sultan {\em et al.}~\cite{sultan2023parallelized}. This allows us to achieve grids of size $L=2^{10},\,2^{11},\,2^{12}$ and $2^{13}$, corresponding to $N=10864,\,42456,\,173824$ and $695296$ cells respectively (see Fig.~\ref{fig:figure1}a). For comparison, {\em Dictyostelium discoideum}, one the largest multicellular organism studied {\em in vivo}, features $80$ to $100$ thousands cells at the end of development, thus seven times smaller than our largest samples. 

\begin{figure}[b]
\includegraphics[width=\columnwidth]{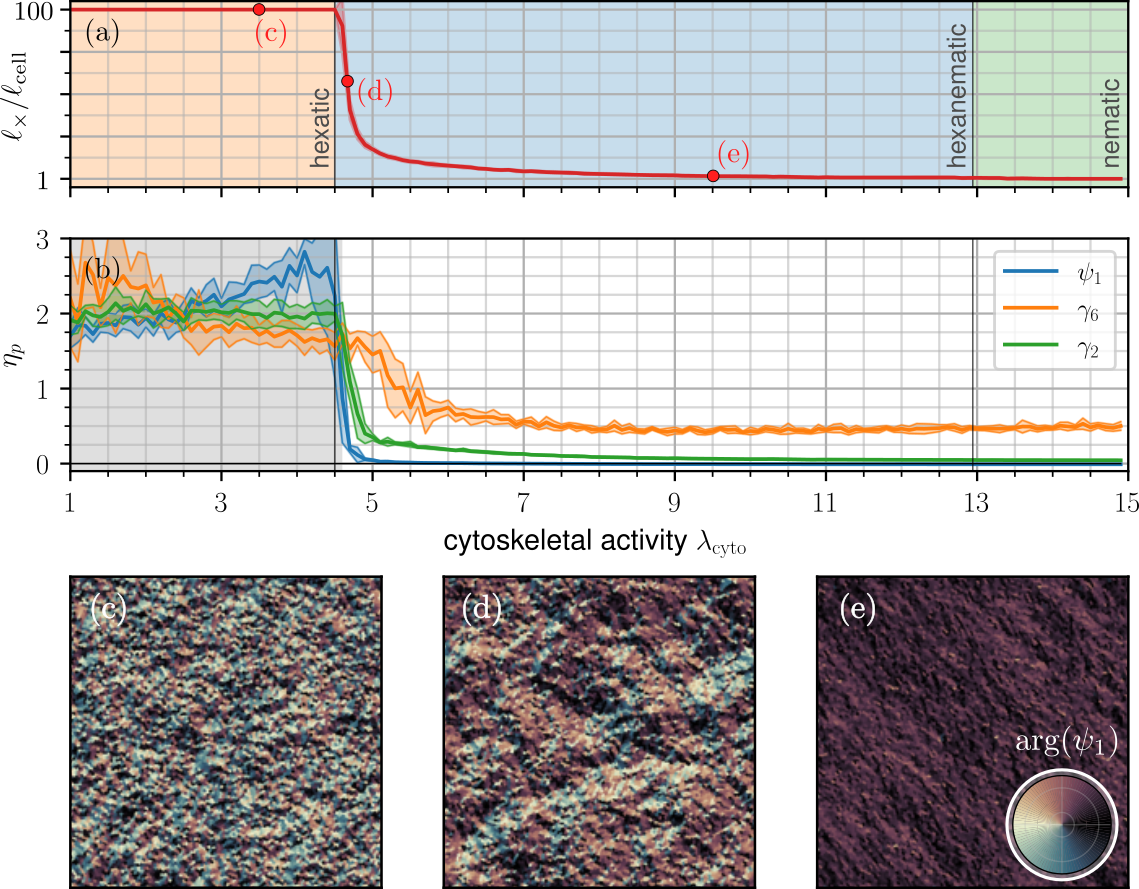}
\caption{
\label{fig:figure3}
(a) Hexanematic crossover length scale $\ell_\times$ and (b) exponents $\eta_{p}$ ($p=1,\,2$ and $6$), for increasing cytoskeletal activity. The red dots mark three specific configurations in the hexatic, hexanematic and nematic phase, whose associated polarization field is displayed in terms of the phase of $\psi_{1}$ in (c), (d) and (e).}
\end{figure}

To explore the phase behavior of our system, we focus on three different orientation fields, hereafter denoted as $\psi_{1}$, $\gamma_{2}$ and $\gamma_{6}$. The former reflects the orientation of the polar axis $\bm{p}=\cos\vartheta\,\bm{e}_{x}+\sin\vartheta\,\bm{e}_{y}$, defined as the position of the actin fibers' center of mass with respect to the cells' centroid and setting the average direction of motion of a cell: i.e. $\psi_{1}=e^{i\vartheta}$~\cite{asano2009pak3}. The latter two quantities, on the other hand, correspond to specific realizations of the generic $p$-atic shape function $\gamma_{p}$ introduced in Ref.~\cite{armengol2023epithelia} and defined as follows. Given a generic {\em irregular} polygon $\mathcal{P}$, whose vertices have position $\bm{r}_{v}$ with respect to the polygon's centroid, $\gamma_{p}=\sum_{v}|\bm{r}_{v}|^{p}e^{ip\arg(\bm{r}_{v})}/\sum_{v}|\bm{r}_{v}|^{p}$, with $p \ge 2$, an integer (Fig.~\ref{fig:figure1}c). The magnitude $|\gamma_{p}|\le 1$ of this complex function, quantifies the resemblance between $\mathcal{P}$ and a {\em regular} $p$-sided polygon having the same size, while its phase $\arg(\gamma_{p})/p$ marks the polygon orientation. Thus, e.g., a regular hexagon centered at the origin of a Cartesian frame and whose vertices have positions $\bm{r}_{v}=\cos(2\pi v/6+\vartheta_{0})\,\bm{e}_{x}+\sin(2\pi v/6+\vartheta_{0})\,\bm{e}_{y}$, with $v=0,\,1\dots 5$, features $|\gamma_{6}|=1$ and $\vartheta=\vartheta_{0}$. The shape function $\gamma_p$ is related to the classical $p$-atic orientational order parameter by $\psi_p = \gamma_p / |\gamma_p|$, and in monodisperse cell layers, it decorrelates with distance at the same rate. In analogy with the orientational order parameter $\Psi_{p}(\ell)=\langle \psi_{p} \rangle_{\ell}\sim\ell^{-\eta_{p}/2}$~\cite{giomi2022long,giomi2022hydrodynamics}, this allows defining the scale-dependent shape parameter $\Gamma_{p}(\ell)=\langle \gamma_{p} \rangle_{\ell}$, such that
\begin{equation}\label{eq:shape_parameter}
\Gamma_{p}(\ell) = \Gamma_{p}(\ell_{\rm cell})\left(\frac{\ell}{\ell_{\rm cell}}\right)^{-\eta_{p}/2}\;,	
\end{equation}
with $\ell_{\rm cell} \approx \sqrt{A_{0}}$ the average cellular size. 

For each simulation, we compute the shape parameters and exponents and classify the corresponding organization based on their relative prominence. The outcome of our parameter space exploration is shown in Fig.~\ref{fig:figure1}d. Phase boundaries are established using an AI-assisted protocol consisting of a first application of the random decision forest classification algorithm (RDFCA)~\cite{ho1995random} followed by a reconstruction of the boundaries via marching cubes~\cite{lorensen1998marching} (see Ref.~\cite{SI} for details). Despite hexatic order being the most prominent in a broad range of material parameters, both nematic and hexanematic order can be detected in our data for sufficiently large cytoskeletal activity~\footnote{Here the hexatetratic phase and more exotic phases were excluded from the visualization (see Ref.~\cite{SI} for details).}. To better highlight the multiscale nature of hexanematic order, in Fig.~\ref{fig:figure2} we show typical configurations of the shape parameters $\Gamma_{2}$ and $\Gamma_{6}$ for two different coarse-graining length scales one order of magnitude apart from each other. Consistently with Eq.~\eqref{eq:shape_parameter}, $p$-atic order decays with the length scale at which is probed (Fig.~\ref{fig:figure2}a). Compared to other liquid crystals with coupled order parameters~\cite{bruinsma1982hexatic,selinger1989tilted,selinger1991dynamics,drouin2022emergent}, however, here the relative prominence of hexatic and nematic order switches across scales, resulting in a substantially different cell organization of the epithelium at the opposite ends of the spectrum. Specifically, hexatic order decays faster than nematic order at any scale $\ell>\ell_{\rm cell}$, but, since $\Gamma_{6}(\ell_{\rm cell})>\Gamma_{2}(\ell_{\rm cell})$, by virtue of the cells' approximatively hexagonal shape, the functions $\Gamma_{2}$ and $\Gamma_{6}$ {\em crossover} at an intermediate scale $\ell_{\times}$ (Fig.~\ref{fig:figure2}b). As a consequence, the emerging cellular patterns feature hexatic (nematic) order at length scales $\ell\ll\ell_{\times}$ ($\ell\gg\ell_{\times}$), in agreement with experiments on MDCK cell monolayers~\cite{armengol2023epithelia,eckert2023hexanematic,armengol2024hydrodynamics}.

To shed light on the origin of this phenomenon, in Fig.~\ref{fig:figure3}a and \ref{fig:figure3}b we plot the crossover length-scale $\ell_{\times}$, together with the exponents $\eta_{p}$ ($p=1,\,2$ and $6$) obtained from the orientational auto-correlation functions $\langle \psi_{p}(\bm{r})\psi_{p}^{*}(\bm{0})\rangle\sim |\bm{r}|^{-\eta_{p}}$. At low $\lambda_{\rm cyto} $ values, $\eta_{6}<\eta_{2}=2$ and orientational order across the epithelium is weakly hexatic. In this regime, increasing $\lambda_{\rm cyto}$ causes a proliferation of T1 events, hence an overall decrease of hexatic order. For $\lambda_{\rm cyto}>4.5$, however, the system transitions to a flocking state, characterized by long-range order among the cells' polar axis. In this regime, $\langle \psi_{1}(\bm{r})\psi_{1}^{*}(\bm{0})\rangle \to {\rm const}$ for $|\bm{r}|\to\infty$, hence $\eta_{1}=0$. Simultaneously, quasi long-range nematic order develops across the system, becoming progressively more prominent as $\lambda_{\rm cyto}$ is increased. Remarkably, the onset of nematic order has the additional effect of restoring hexatic order, but only at length scales that becomes smaller and smaller as cytoskeletal activity is increased, thereby giving rise to hexanematic order. In this regime, $\eta_{2}<\eta_{6}<2$ and nematic order undergoes a surprisingly abrupt enhancement, becoming first quasi long-range ($\eta_{2}<1/4$ at $\lambda_{\rm cyto}\approx 5$) and eventually long-range ($\eta_{2} = 0$ at $\lambda_{\rm cyto} \approx 13$). Notice that the exponents $\eta_{1}$, $\eta_{2}$ and $\eta_{6}$ crosses at the critical point, thus indicating the possibility of a causal relation between hexanematic order and the onset of flocking. Finally, for large $\lambda_{\rm cyto}$ values, cells organize in a single ``flock'' translating across the plane at constant speed while $\eta_{6} \approx 1$. In this regime, increasing the cytoskeletal activity no longer affects the frequency of T1 events in the nonequilibrium steady state, nor the amount of hexatic order (Fig.~\ref{fig:figure3}b).

\begin{figure}[t]
\includegraphics[width=\columnwidth]{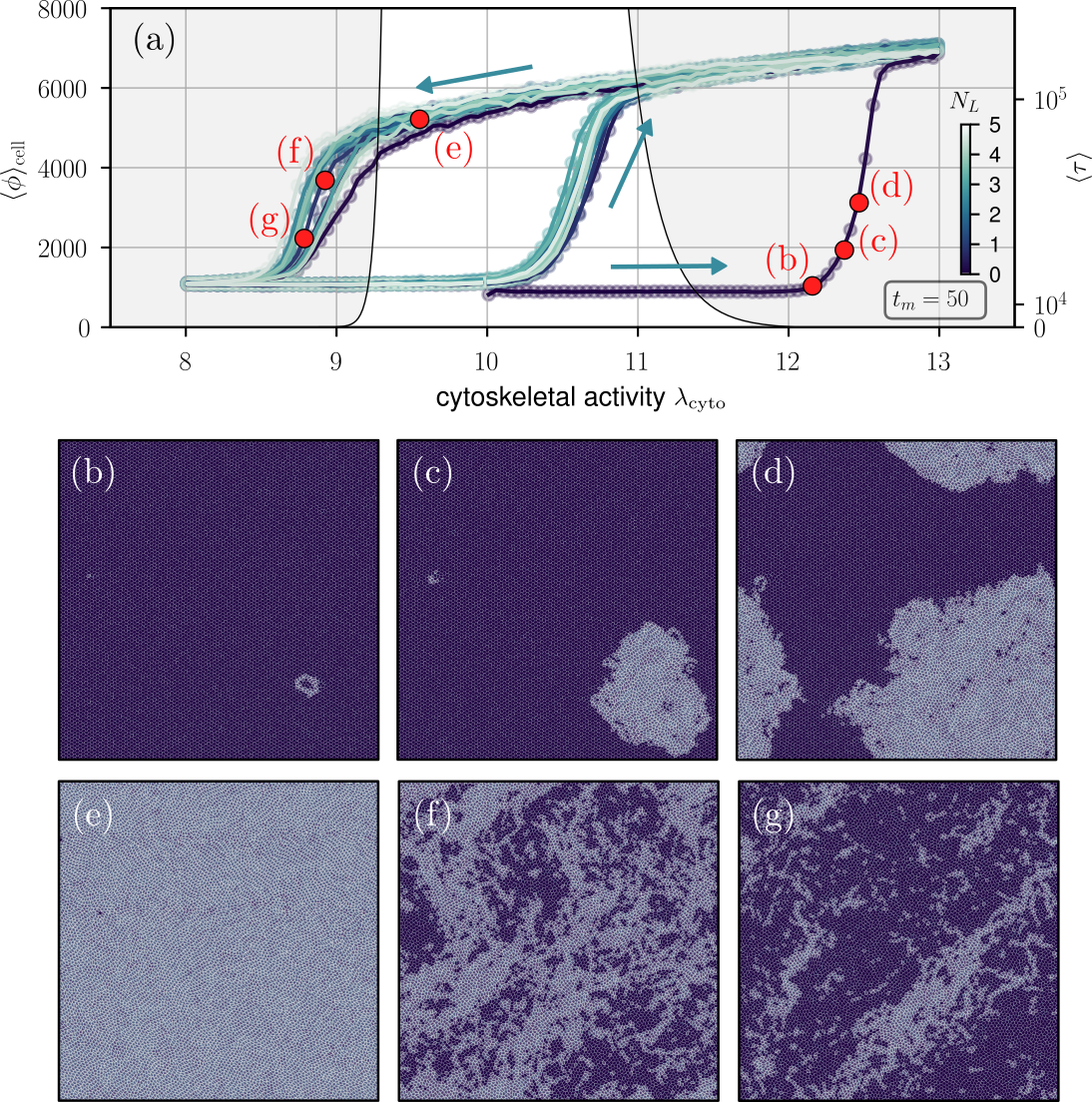}
\caption{
    \label{fig:figure4}
    Hysteresis in the cellular Potts model. 
    In (a) the hysteresis loop is shown for the \textit{mean actin per cell} $\langle \phi \rangle_{\rm cell}$ as a function of the cytoskeletal activity parameter $\lambda_{\rm actin}$ on the left $y$-axis. Here a single system configuration travels along the loop as snapshots are taken every $t_m$ MCS. Size $N=1024$ is used, with $T=3$. On the right $y$-axis the \textit{average time of transition} $\langle \tau \rangle$ is shown as a function of $\lambda_{\rm actin}$. For the curve on the left this is from motile to non-motile, and for curve on the right is from non-motile honeycomb to motile cells. $N_L$ is the number of loops. In (b,c,d) consecutive snapshots show a spreading front of collectively migrating cells originating from a single fluctuation of cytoskeletal activity, bringing the system in a motile state. In (e,f,g) consecutive snapshots show motile cells that cease collective migration into a non-motile state as $\lambda_{\rm cyto}$ is decreased. 
}
\end{figure}

In the above analysis of MsO in migrating cell monolayers, noise is sufficiently high to guarantee intercalation at any finite $\lambda_{\rm cyto}$ value, thus preventing cells from crystallizing. For $\lambda_{\rm cyto}=0$, on the other hand, decreasing $T$ yields crystalline epithelia characterized by long-range $6$-fold orientational order, consistent with previous studies of the CPM~\cite{durand2019thermally,nemati2024cellular} and other cell-resolved models of epithelia~\cite{li2018role,pasupalak2020hexatic,li2021melting}. Unlike these models, however, explicitly accounting for cytoskeletal activity unveils a striking example of phenotypic hysteresis, obtained upon varying $\lambda_{\rm cyto}$ at low $T$ values. Fig.~\ref{fig:figure4} shows a typical hysteresis loop obtained upon initializing the system in a honeycomb lattice at $T=3$ and cycling $\lambda_{\rm cyto}$. A phenotypic coordinate can be obtained from the mean actin density per cell: i.e. $\langle \phi \rangle_{\rm cell}$. This is plotted against $\lambda_{\rm cyto}$ in Fig.~\ref{fig:figure4}a, while Figs.~\ref{fig:figure4}b-g show the typical configurations observed along the loop. In all panels, background colors reflects the local actin density, so that light and dark tones mark the position of motile and non-motile cells respectively.  

In conclusion, our large-scale numerical simulations augmented with a simple but explicit representation of actin polymerization, reveal a complex yet well-structured organization across length scales. This organization emerges from the dynamical interplay between shape-mediated intercellular interactions and cell motility. Although it has no counterpart in thermal systems, several of its features can be interpreted using the language of liquid crystals and orientational order, with anomalous dimensions $\eta_p$ extending beyond the bounds imposed by equilibrium statistical physics. Furthermore, when noise is sufficiently low to permit crystallization at finite actin polymerization rate, cycling the monolayer across the melting transition produces a clear example of phenotypic hysteresis. Exploring this phenomenon further may provide a route toward a quantitative understanding of phenotypic plasticity across the epithelial–mesenchymal transition (EMT)~\cite{friedl2010plasticity}. In this context, mesenchymal cells can retain their phenotype even after EMT-inducing signals, such as TGF-$\beta$, are withdrawn~\cite{youssef2024two}. In cancer, this persistent state is associated with enhanced metastatic potential, making hysteresis a clinically relevant indicator of poor prognosis~\cite{jahun2023leaked}. A complete understanding of this behavior will require incorporating in our simulations biochemical pathways beyond the actin polymerization–depolymerization cycle; a promising framework for future studies of phenotypic plasticity.

\begin{acknowledgments}
This work is partially supported by the ERC-CoG grant HexaTissue (C.C.B. and L.G.). 
\end{acknowledgments}

\bibliography{cpm_2026_05_13.bbl}

\end{document}


\title{Multiscale order, flocking and phenotypic hysteresis\\in the cellular Potts model of epithelia:\\Supplementary information}

\author{Calvin C. Bakker}
\affiliation{Instituut-Lorentz, Leiden Institute of Physics, Universiteit Leiden, P.O. Box 9506, 2300 RA Leiden, The Netherlands}
\author{Marc Durand}
\affiliation{Universit\'{e} Paris Cit\'{e}, CNRS, Mati\`{e}re et Syst\`{e}mes Complexes, F-75013 Paris, France}
\author{Fran\c{c}ois Graner}
\affiliation{Universit\'{e} Paris Cit\'{e}, CNRS, Mati\`{e}re et Syst\`{e}mes Complexes, F-75013 Paris, France}
\author{Luca Giomi}
\affiliation{Instituut-Lorentz, Leiden Institute of Physics, Universiteit Leiden, P.O. Box 9506, 2300 RA Leiden, The Netherlands}
\email{giomi@lorentz.leidenuniv.nl}

\maketitle

\section{Cellular Potts Model Implementation for GPU}
\label{sec:cpm}
The cellular Potts model (CPM) is implemented in the CUDA parallel programming language~\luebke{}, building on the work of Sultan \emph{et al.}~\sultan{}. Using the open-source code published by them, the algorithm was forked and extended by incorporating the fragmentation-free method of Durand and Guesnet~\duranda{}. Each simulation uses a cell-area constraint set to $A_0 = 96.51$, where this value was found to be a close-to-optimal solution of the optimization problem to fit an unfrustrated honeycomb packing onto a square grid of size $L = 2^{10}$. Following the parameters in Ref.~\durandb{}, the adhesion parameter is set to $J = 16.0$. We simulate systems of size $L=2^{10}$ on NVIDIA T600 and RTX 2080ti graphical processing units and of sizes $L=2^{11}$, $2^{12}$ and $2^{13}$ on NVIDIA RTX 4090.

\section{Open-source code}
The code used to produce the cellular Potts model simulations will be open-sourced upon publication of the manuscript, and is available on Github as \texttt{cpm\_nano}: ``A parallel implementation of the two-dimensional cellular Potts model in CUDA'', and the code used to compute measures such as the order parameters will be open-sourced as \texttt{statphys2D\_cuda}: ``A collection of statistical physics tools for analysis of two-dimensional systems'' both on on \url{github.com/softbiomech/}. The code will be made available under the GNU GPLv3 license, ensuring that users have the freedom to use, modify, and redistribute the software.

\section{Parameter space exploration}

To explore what phases can be produced by the CPM Hamiltonian given in the main text, a three-dimensional parameter space is chosen in which the cytoskeletal activity parameters and the \red{noise introducing fluctuation allowance parameter} are varied. Geometric parameters are kept constant. The three parameters are the cytoskeletal activity parameters $\lambda_\text{cyto}$ and $\Phi$, and the fluctuation allowance parameter $T$. For setting the geometric parameters the values are used which are specified in Sec.~\ref{sec:cpm}, while for the other parameters the ranges are $\lambda_\text{cyto} \in [0,25)$, $\Phi \in [0,250)$, and $T \in [0,30)$.

The dataset consists of 901 CPM configurations of size $L=1024$ propagated for 100.000 \red{MCS}. We will mention this dataset as the parameter space exploration (PSE) dataset. 
The parameter values $(T,\lambda_\text{cyto},\Phi)$ were obtained through a Monte Carlo method (random uniform sampling), where for each sample the CPM is initialized as a honeycomb lattice of cells with periodic boundary conditions as the initial condition. To avoid sampling two instances of parameters that are too similar the distance to all other points in computed and the sample is rejected when this distance is below a given threshold. For each of the configurations $\Omega$ obtained from a random sampling of $(T,\lambda_\text{cyto},\Phi)$ the corresponding order parameters can be computed, which span a three-dimensional phase diagram. For our purposes the most insightful classifications is the \textit{ordered numbering of multiscale orientational order}, which represents multiscale order by decimal strings. Here the order is indicative of the scales from smallest to largest. For example, hexatic order over all scales would be class `$6$', nematic order over all scales would be `$2$', hexanematic order would be `$62$' (small scale hexatic, large scale nematic), nemahexatic order would be `$26$', and hexanematetratic order would be `$624$'. Following the same naming convention as used for hexanematic order (hexatic-nematic), this stands for hexatic-nematic-tetratic. Meaning that on the smallest scale the hexatic order is dominant, on intermediate scales the nematic order, and on the largest scale the tetratic order. 

In Fig. 1d in the main text the PSE dataset is represented by visualizing the boundary mesh of the regions belonging to the different types of order that are present. This boundary mesh is created by fitting a random decision forest classification algorithm (RDFCA)~\ho{} to the dataset, and then predicting the boundary by the marching cubes algorithm (MCA)~\lorensen{}. First we will explain the RDFCA, and then we will explain how the boundary mesh is created from the predicted probabilities of the RDFCA by the MCA.

The RDFCA was configured to use $200$ estimators, the Gini impurity criterion, and data bootstrapping. We use the \texttt{scikit-learn} implementation for this purpose~\pedregosa{}. The algorithm creates multiple decision trees (a ``forest''), where each tree is trained on a random subset of the data and a random subset of the features, and the final prediction is made by averaging the predictions of all the trees. The Gini impurity criterion and data bootstrapping are used to reduce overfitting and improve the generalization of the model. The features the algorihm uses are the parameters $(T,\lambda_\text{cyto},\Phi)$, and the target variable for prediction is the ordered numbering of multiscale orientational order (e.g. `62'). For verifying that the RDFCA is capable to produce a good representation of the PSE dataset, it was tested to have an accuracy of $87.74\pm2.14\%$ on 80-20 splits of the dataset (of 80\% fitting and 20\% testing) for 50 random initializations, where the best found fit of $91.16\%$ was used to create the boundary mesh as seen in Fig. 1d in the main text. In this picture the phase `$64$' is omitted from the figure for a more clear and insightful representation, as this phase spans almost the whole high-\red{fluctuation allowance} regime and would mask the boundaries of the other phases visually. Other omitted phases are `$24$', `$624$', and `$264$' which make up only $2.66\%$ of the dataset. They are present in the high-fluctuation allowance high-activity regime.

\begin{figure*}
\includegraphics[width=0.7\textwidth]{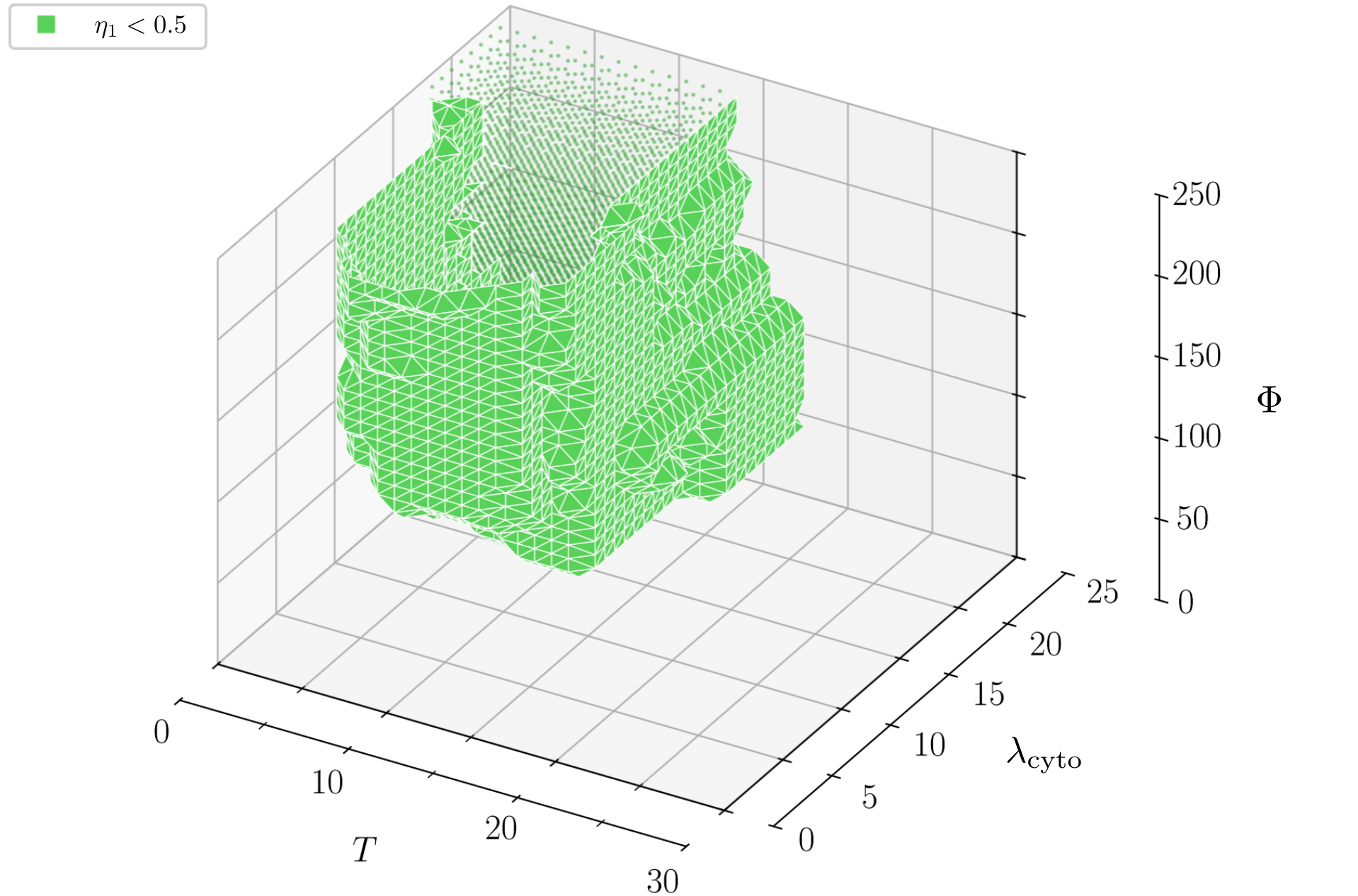}
\caption{\label{fig:slice} 
Three-dimensional slice of the parameter space where the RDFCA predicted the phase boundary for long-range order (LRO) in the polarization order parameter $\psi_1$. The LRO is quantified by computing the correlation function for $\psi_1$, performing a power-law fit for the correlation function, and then computing the $\eta_1$ exponent from the power-law fit. The phase boundary is determined by the value of $\eta_1$ being below a certain threshold (set to $\eta_1<0.5$ here). The surface shown in this figure represents this boundary, which is consistent with the boundary predicted for the onset of multiscale orientational order as shown in Fig. 1b in the main text, indicating the connection between the onset of multiscale orientational order and LRO in the cell-polarization.
}
\end{figure*}

We used the MCA to create a boundary mesh from the predicted probabilities of the RDFCA. The MCA is a computer graphics algorithm that extracts a polygonal mesh of an isosurface from a three-dimensional scalar field. In our case, the scalar field is given by the predicted probabilities of the RDFCA for each class of order, where the value at each point in the parameter space corresponds to the probability of that point belonging to a certain class. The threshold for the MCA is set to zero, meaning that the algorithm will extract the boundary where the predicted probability of a class changes from being the most likely class to being less likely than another class. This results in a polygonal mesh that represents the boundary between different classes of order in the parameter space, which can be visualized in Fig. 1d in the main text. For our implementation of the MCA we used the \texttt{scikit-image} implementation~\vanderwalt{}. The MCA works by iterating through the scalar field and identifying the points where the value crosses the threshold. It then creates a polygonal mesh by connecting these points, which results in a representation of the isosurface. The mesh allows us to visualize the different regions of order in the parameter space in three dimensions. 

In addition to the figure given in the main text in Fig. 1d, a three-dimensional slice is shown of the parameter space where the RDFCA predicted the phase boundary where we have long-range order (LRO) in the polarization order parameter $\psi_1$. This LRO is quantified by computing the correlation function for the order parameter, performing a power-law fit for the correlation function, and then computing the $\eta_1$ exponent from the power-law fit. The phase boundary is then determined by the value of $\eta_1$ being below a certain threshold (set to $\eta<0.5$ here). In Fig. \ref{fig:slice} the boundary is shown as a surface. What we can observe from this figure is that this boundary is consistent with the boundary predicted for the onset of multiscale orientational order as shown in Fig. 1b in the main text, which indicates the connection between the onset of multiscale orientational order and long-range polarization alignment. To quantify the connection between the hexanematic phase and the LRO in the polarity (the flocking phase) we computed the overlap between different types of multiscale orientational order and the presence of LRO in the polarization. The found percentages for LRO in the polarization are $3.83\%$ for the hexatic phase, $95.77\%$ for the hexanematic phase, $93.67\%$ for the nematic phase, and $0.00\%$ for the high-noise hexatetratic phase. These percentages are obtained from the PSE dataset.

\section{Computation and coarse graining of the shape function $\gamma_p$}

To compute the shape function $\gamma_{p}$, we construct a polygonal tiling of a configuration generated by the CPM. In confluent cell layers, this tiling is unique due to the absence of gaps between the cells. For this reason we need the fragmentation-free algorithm. The polygonization is obtained by identifying intersection points where three or more cells meet, and placing the vertex of a polygon at this intersection point. These points are topologically well-defined due to the confluency. The resulting polygon vertices are then used to compute $\gamma_{p}$ for each individual cell in the configuration $\Omega$.

A coarse graining of the $\gamma_p$ order parameter is performed by local averaging of the order parameter in space, where the radius of coarse graining $\ell$ can be varied to quantify how the order parameter is distributed over multiple length scales. The algorithmic implementation of the procedure is given in the CUDA library \texttt{statphys2D\_cuda}, making use of GPU parallelism to efficiently perform the coarse graining. The procedure starts with specifying a subgrid of size $(M,M)$, where $M<N$, and searching for each subgrid point $i$ the set of cells whose center-of-mass position $\vec{r}_n$ lies within a distance $\ell$ from that subgrid point. Denoting this neighborhood by $\mathcal{N}_i(\ell)$, the coarse-grained value at subgrid point $i$ is computed by averaging over the cells in this neighborhood,
\begin{equation}
\bar\gamma_p^i(\ell)=\left\langle \gamma_p(n) \right\rangle_{n\in\mathcal{N}_i(\ell)}.
\end{equation}
The next step is to take the average of $\bar\gamma_p^i(\ell)$ over all points of the subgrid. The resulting quantity will be denoted by $\Gamma_p(\ell)$ with the mathematical expression $\Gamma_p(\ell) = \langle \bar\gamma_p^i(\ell) \rangle_{i}$, following J. Armengol-Collado \emph{et al.}~\collado{}. The absolute value $|\Gamma_p(\ell)|$ is interpreted as quantifying the $p$-atic order of the configuration $\Omega$ at length scale $\ell$. 

\begin{figure}
\includegraphics[width=0.7\textwidth]{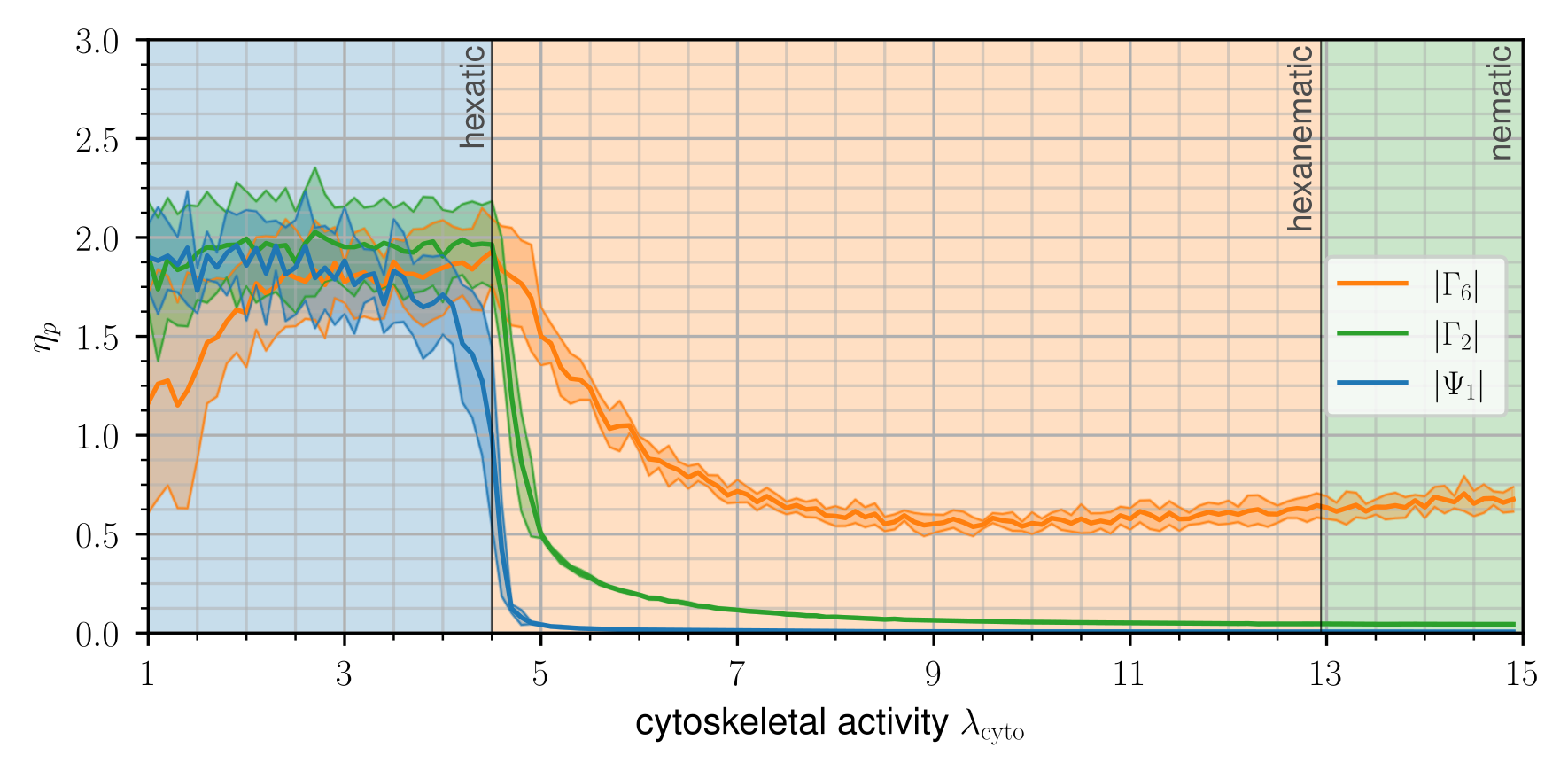}
\caption{\label{fig:eta_coarsegrain} 
\red{Power-law fit exponent $\eta_p$ extracted from the coarse-grained order parameter $|\Psi_1|$ obtained from the polarization $\psi_1$ and the coarse-grained order parameter of $|\Gamma_p|$ for $p=2$ and $p=6$, shown as a function of the cytoskeletal activity parameter $\lambda_{\rm cyto}$. The exponent is obtained by computing $|\Psi_1(\ell)|$ for multiple coarse-graining radii $\ell$ and fitting its scaling with $\ell$ to a power law, $|\Psi_1(\ell)|\sim \ell^{-\eta_p/2}$. Error bars indicate one standard error of the mean, obtained from multiple simulations.}
}
\end{figure}

In previous work it was shown that hexanematic order can be found in epithelial tissues by analyzing the multiscale order as measured by the $|\Gamma_p|$ order parameter~\collado{}. Hexanematic order is quantified by comparing two types of liquid crystal order (LCO). These two types of LCO are hexatic and nematic orientational order, given by $p=6$ and $p=2$ respectively. Comparing these for a given configuration, the signature of hexanematic order is a crossover at a certain length scale $\ell_\times$. At the smaller length scales the hexatic order parameter is larger than the nematic order parameter, while after the crossover the nematic order parameter is larger. Following the work of Ref~\collado{} the scaling of $|\Gamma(\ell)|$ is quantified by fitting to a power law as $|\Gamma(\ell)| \sim \ell^{-\eta_p/2}$, where $p=\{6,2\}$ in our analysis; resulting in the exponents $\eta_6$ and $\eta_2$. 

\red{
In Fig.~\ref{fig:eta_coarsegrain}, for completeness in relation to Fig.~3, the exponent $\eta_p$ as a function of the cytoskeletal activity parameter $\lambda_{\rm cyto}$ is given for the coarse-grained order parameter $|\Psi_1(\ell)|$ obtained from the polarization $\psi_1$ and the coarse-grained order parameters of $|\Gamma_p|$ for $p=2$ and $p=6$. Other parameters are fixed at $T=10$ and $\Phi=145$. The exponent is obtained by computing $|\Psi_1(\ell)|$ for multiple coarse-graining radii $\ell$ and fitting its scaling with $\ell$ to a power law.
}

\red{
Reproducing hexanematic order in numerical simulations is non-trivial. In previous work it was shown that the self-propelled Voronoi (SPV) model~\bi{} was not able to produce hexanematic order, which seems to be caused by property of the model that cells are represented as polygons~\collado{}. As cell shapes are restricted to polygonal shapes, they might not have the necessary amount of degrees of freedom needed for hexanematic order to emerge. In contrast, the multi-phase field (MPF) model~\loewe{} and CPM were found to be able to produce hexanematic order, as they have more degrees of freedom to shape the cell, which in turn allows for more diverse and realistic cell shapes. The MPF model is a good candidate to study hexanematic order numerically, but due to the computational resources needed to simulate the coupled partial differential equations of the model, simulating cell monolayers with hundreds of thousands of cells is not feasible. Therefore it is not possible to scale-up to very large systems. From this point of view the CPM is an alternative model to use which is similar to the MPF model, as they both simulate cells as cell contour fluctuations with physically realistic details, in contrast to the SPV and vertex models~\nagai{} in which cells are modelled as polygons. The CPM has the advantage of doing so more efficient than the MPF model when considering the computational resources needed per cell; even more so since the introduction of the algorithm by S. Sultan \emph{et al.}~\sultan{}.
}

\section{Hysteresis}
\begin{figure}
\includegraphics[width=0.6\textwidth]{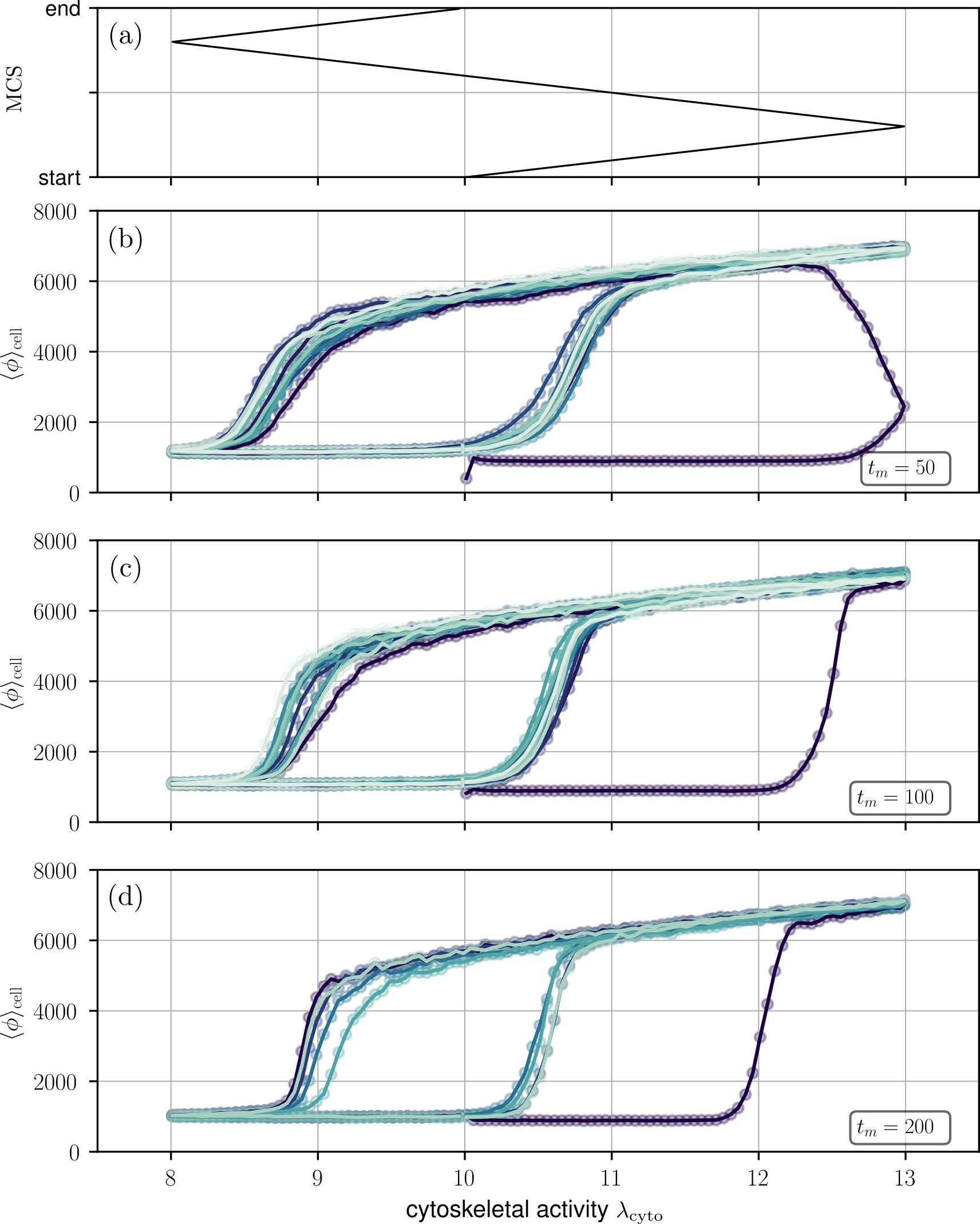}
\caption{\label{fig:hystloops}
Hysteresis loops with the loop schedule shown in (a), for multiple different values of the sampling time $t_m$ in (b-d). The hysteresis loops are obtained by initializing the cellular Potts model in a honeycomb configuration, and then making $5$ loops following the given schedule while keeping the other parameters fixed. For each value of $\lambda_\text{cyto}$ the system is simulated for $t_m$ \red{MCS}, where the order parameter is measured at the end of this time interval. 
}
\end{figure}

\begin{figure}
\includegraphics[width=0.99\textwidth]{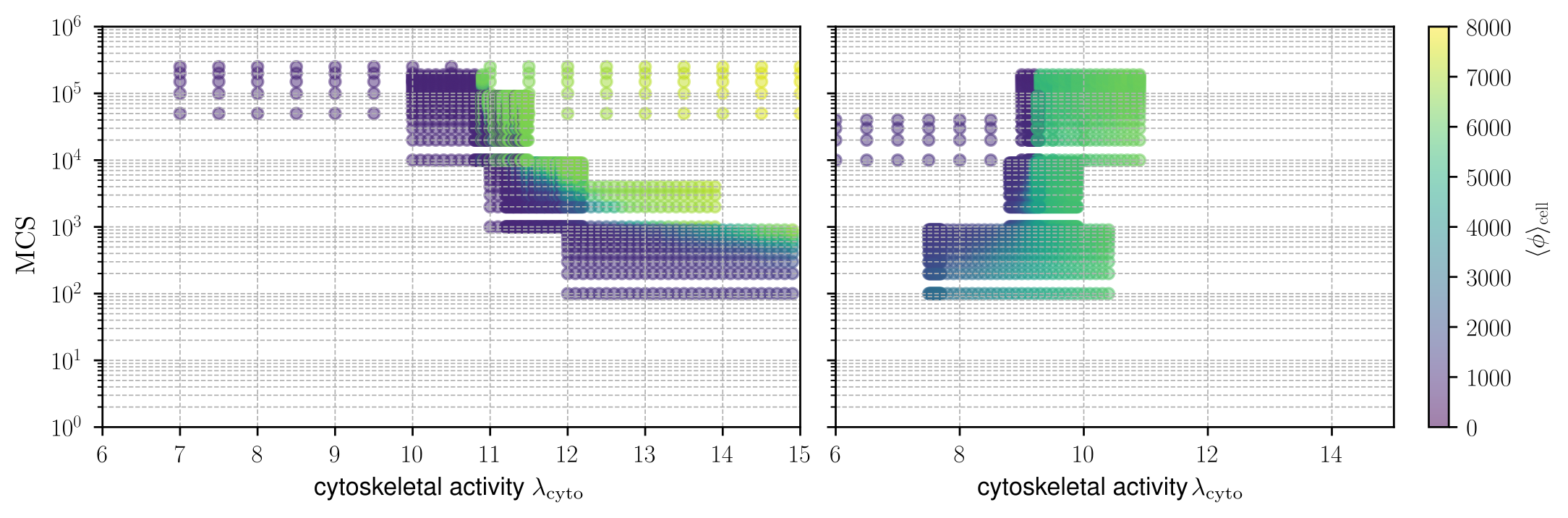}
\caption{\label{fig:transition_prob} Transition probabilities for the transitions from honeycomb configuration to liquid (left), and from liquid to honeycomb configuration (right), as a function of the cytoskeletal activity parameter $\lambda_\text{cyto}$.
}
\end{figure}

To investigate the presence of hysteresis in the \red{cytoskeletal activity parameter $\lambda_\text{cyto}$} we perform a series of \red{Cellular Potts Model} simulations where the system is initialized in a honeycomb configuration, where crucially the \red{fluctuation allowance} is set to $T=3$, in contrast to other simulaitons performed for this work. For each value of $\lambda_\text{cyto}$, the system is simulated for a fixed number of \red{MCS} $t_m$, and the \textit{average actin per cell} $\langle \phi \rangle_{\rm cell}$ is measured at the end of this time interval. The schedule of $\lambda_\text{cyto}$ is shown in \red{Fig. \ref{fig:hystloops}a}, where the value of $\lambda_\text{cyto}$ is varied between the values $8.0$ and $15.0$ for $5$ loops. The resulting hysteresis loops are shown in \red{Fig. \ref{fig:hystloops}b-d} for different values of the sampling time $t_m$. 
To quantify this effect in a trajectory-independent way the transition probabilities for both transitions are computed as a function of $\lambda_\text{cyto}$, where the transition from the honeycomb configuration to the liquid configuration is shown in Fig. \ref{fig:transition_prob}a, and the transition from the liquid configuration back to the honeycomb configuration is shown in Fig. \ref{fig:transition_prob}b. From these figures it can be observed that the transition from the honeycomb configuration to the liquid configuration occurs at a lower value of $\lambda_\text{cyto}$ than the transition from the liquid configuration back to the honeycomb configuration, which is indicative of hysteresis. 

\section{Static structure factor $S(q)$}

\begin{figure}
\includegraphics[width=0.9\textwidth]{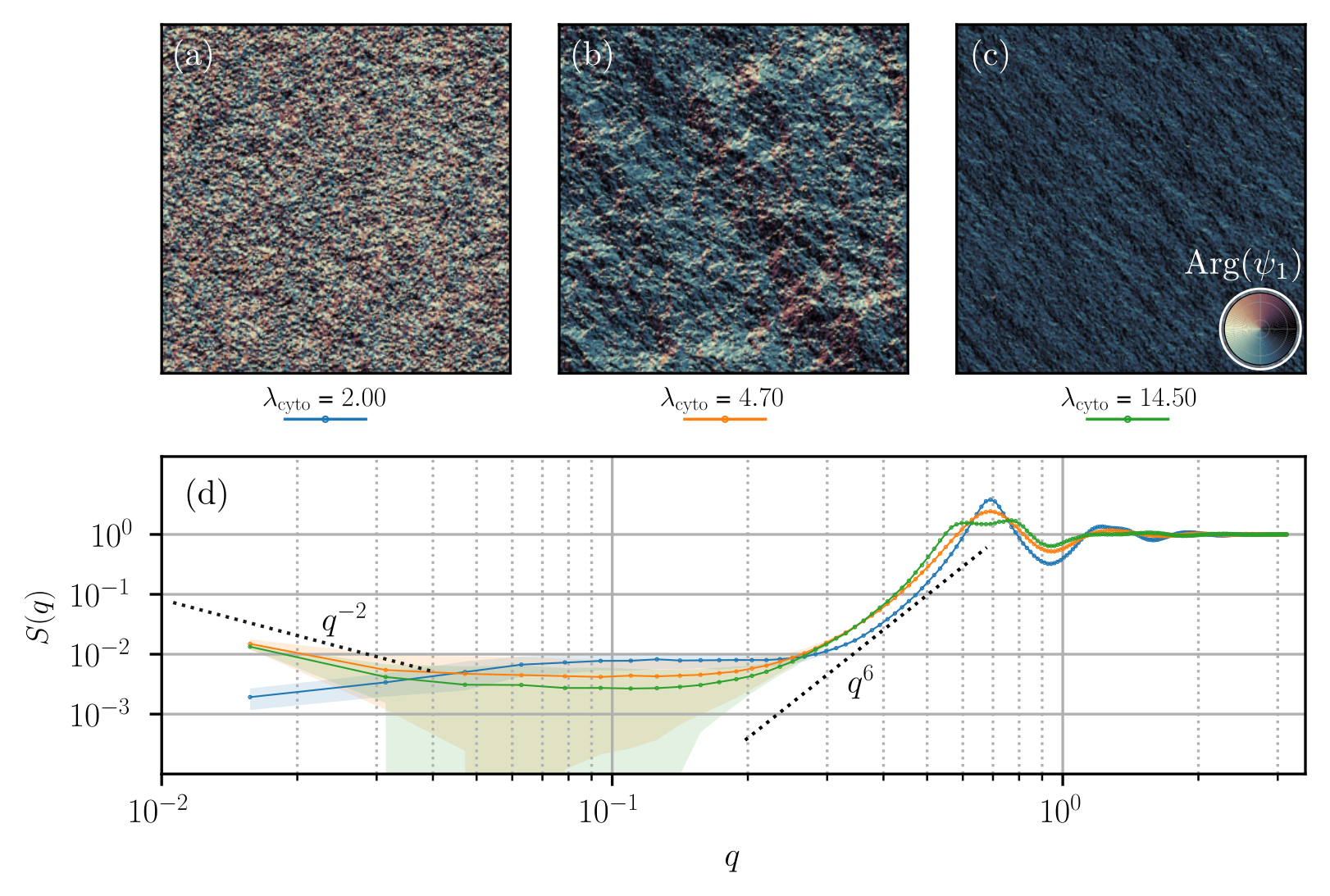}
\caption{\label{fig:ssf} 
\red{The polarization and radial static structure factor as a function of $\lambda_\text{cyto}$. In (a-c) a CPM configuration is shown per panel for $L=2048$, where cells are colored based on the orientation of their polarization $\psi_1$, going from low $\lambda_\text{cyto}$ in (a) with dominant hexatic order, intermediate in (b) with hexanematic order, and high in (c) with nematic order. Here one can clearly observe the flocking structures based on the polarity of the cells. In (d) the radial structure factor $S(q)$ is shown, averaged over multiple configurations of the same $\lambda_\text{cyto}$ value.}
}
\end{figure}

\red{
Computing the static structure factor $S(q)$, and measuring the value $\lim_{q\rightarrow 0}S(q)$, we have a complementary measure with respect to $|\Gamma_p(\ell)|$ to see if large scale structure is present in the configuration $\Omega$. The computation of the static structure factor is based on the formulae presented in the notes of Zhang~\zhang{}, where for the CPM the center-of-mass coordinates $\vec{r}_i$ are used as the density profile $\rho$ in the formalism to compute $S(q)$. From \red{Fig.}~\ref{fig:ssf} we can observe that for low values of $\lambda_\text{cyto}$ the value of $S(q)$ at small $q$ is close to decaying, indicating the absence of large scale structure, while for higher values of $\lambda_\text{cyto}$ the value of $S(q)$ at small $q$ is trending upwards, indicating the presence of large scale structure. From the data it can be observed that although in this paper we consider hexanematic order to be \emph{multiscale} and nematic order not, both these phases produce a radial static structure factor that grows as we take the limit $\lim_{q\rightarrow 0}S(q)$, which seems to be proportional to the same power of $q$ for both. The largest variation for both is in the intermediate $q$ regime where $q\in(2\cdot 10^{-2},1\cdot 10^{-1})$.}

\section{Finite size scaling}
\begin{figure}
\includegraphics[width=0.9\textwidth]{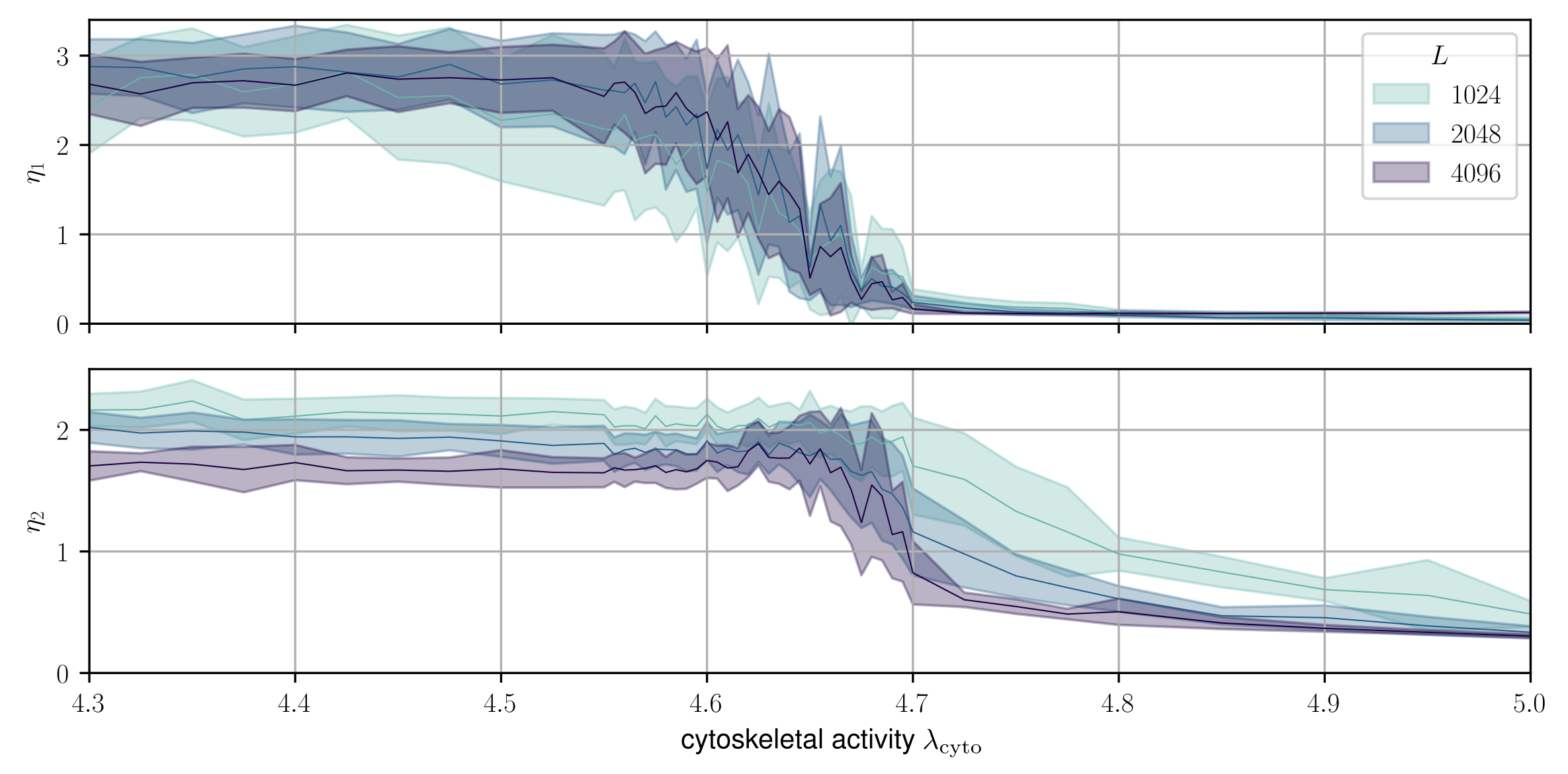}
\caption{\label{fig:finitesize} 
Finite size scaling of the transition from hexatic to hexanematic order. \red{As the system size increases, }the transition becomes sharper for the nematic exponent $\eta_2$, not for the polarization exponent $\eta_1$.}
\end{figure}

To investigate the finite size scaling of the transition from hexatic to hexanematic order and the onset of LRO, we perform a series of simulations where the system size $L$ is varied while keeping the other parameters fixed. The activity parameter $\lambda_\text{cyto}$ is varied in the range $\left[4.3,5.0\right)$. For each value of $L$, the system is simulated for a fixed number of MCS, and the order parameters are measured at the end of this time interval. The resulting data is shown in Fig. \ref{fig:finitesize} for the exponents $\eta_1$ and $\eta_2$ obtained from the correlation functions, where it can be observed that as the system size increases, the transition from hexatic to hexanematic order becomes sharper for the nematic exponent $\eta_2$, while for the polarization exponent $\eta_1$ the transition does not become sharper with increased system size.

